\documentclass[12pt]{iopart}

\usepackage{graphicx} 
\usepackage{colordvi}
\usepackage[normalem]{ulem}
\begin{document}

\title[A twisted geometry in non-central Pb+Pb collisions]{A twisted emission
geometry in non-central Pb+Pb collisions measurable via azimuthally sensitive HBT}

\author{Gunnar Graef $^{1,2}$, Mike Lisa $^{3}$, Marcus Bleicher $^{1,2}$}
\address{$^1$ Institut f\"ur Theoretische Physik, Goethe Universit\"at
Frankfurt,Germany} 
\address{$^2$ Frankfurt Institute for Advanced Studies
(FIAS), Ruth-Moufang-Str. 1, 60438 Frankfurt, Germany}
\address{$^3$ Department of Physics, Ohio State University, Columbus, Ohio
43210, USA}

\ead{graef@th.physik.uni-frankfurt.de}

\begin{abstract}
We use the Ultrarelativistic Quantum Molecular Dynamics (UrQMD) model to
simulate Pb+Pb collisions. In the freeze out geometry of non-central Pb+Pb
collisions we observe a tilt of the particle emission zone in the collision
plane away from the beam axis. We find that the magnitude of this tilt depends on the scale at which the distribution is measured. We quantify this ``twisting'' behavior with a parameterization and
propose to measure it experimentally via azimuthally sensitive
Hanbury-Brown Twiss correlations. Additionally we show that the twist is
related to the emission of particles from different times during the evolution
of the source. A systematic comparison between the theoretically observed
twist in the freeze out position distribution and a mock experimental
analysis of the model calculations via HBT correlations is shown.
\end{abstract} \maketitle

\section{Introduction}

Quantum-Chromo-Dynamics (QCD) is the theory that describes the properties of
strongly interacting matter. Currently only a few problems in QCD can be solved
analytically. To explore the details of QCD experimentally one needs to
compress and heat up QCD matter to regimes that were present microseconds after
the Big Bang. Today similar conditions are present only in the interior of
neutron stars or in heavy-ion collisions at relativistic energies. Several
experimental programs at the SPS (e.g. NA49, CERES and NA50/NA60), RHIC (e.g.
PHENIX, STAR, PHOBOS and BRAHMS) and at the LHC (e.g. ALICE, CMS and ATLAS) are
dedicated to study the medium created in heavy-ion collisions. Due to the small
scale and short lifetime of the reactions, only the momentum spectrum of the
particles coming out of the interaction zone can be measured directly. However,
the space-time structure of the collisions can be probed indirectly using
Hanbury-Brown Twiss (HBT) interferometry. This technique uses two particle
correlations to probe the space-time shape of the particle emission zone of
relativistic heavy ion collisions. Extensive studies have been performed
\cite{Lisa:2005dd} to map out the dependence on transverse momentum ($p_T$),
rapidity ($y$), collision energy ($\sqrt{s}$) and charged particle
multiplicity. HBT interferometry measurements relative to the impact parameter
direction yield additional insights about the shape and orientation of the
emission region as a whole. Unfortunately only a small number of azimuthally
sensitive HBT measurements have been reported so far
\cite{Lisa:2000xj,Adams:2003ra,Adamova:2008hs}. However the beam energy scan
initiative (RHIC-BES) at the Relativistic Heavy Ion Collider (RHIC) and the
CBM-Experiment at FAIR will bring new measurements of this observable
\cite{Aggarwal:2010cw,Anson:2011ik} over a broad energy range.

In section \ref{sec:freeze} of this paper we explain how to extract the
substructure of the tilt of the pion freeze out distribution in the event plane
from the spatial freeze out distribution. We discuss a new feature of the
source, the twist, not yet measured or discussed extensively in the literature.
A parameterization of the twist and aspects of its physical origin are
discussed. Section \ref{sec:asHBT} suggests a phenomenological approach that
allows to measure the twist experimentally and an example of the results from
such an approach.

\section{Analysis of the pion freeze out distribution \label{sec:freeze}}
We use the well known Ultrarelativistic quantum Molecular Dynamics (UrQMD)
approach to simulate the heavy-ion events for this paper. UrQMD is a hadronic
non-equilibrium transport approach that produces particles via string
fragmentation and resonance excitation and decay. For details of UrQMD the
reader is referred to \cite{Bleicher:1999xi,Bass:1998ca,Li:2006gp}. Previous
HBT results of the model can be found in \cite{Li:2007yd,Li:2008qm,Li:2012ta}

The anisotropic ``almond'' shape of the emission zone in the transverse plane
created in non-central collisions is discussed extensively in the literature,
as it leads to momentum-space anisotropies (elliptic
flow)~\cite{Ollitrault:1992bk}. However, the spatial substructure of the
emission region is much richer than its projection onto the transverse plane.
The projection onto the reaction plane, $x-z$ (where $x$ is the direction of
the impact parameter and $z$ is the beam direction)-- {\it even when selecting
particles emitted only at midrapidity}-- is characterized by a nontrivial shape
and anisotropies.

\begin{figure}[ht!]
\centering
\includegraphics[width=0.4\textwidth]{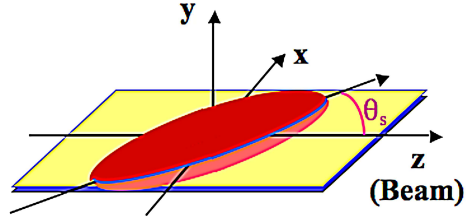}
\caption{Orientation of the particle emitting zone (red) in the reaction plane and
definition of the tilt angle $\theta_S$. Taken from \cite{Mount:2010ey}
\label{fig:TiltCartoon}}
\end{figure}

Transport calculations and three-dimensional hydrodynamic simulations generate
distributions characterized by a tilt relative to the beam direction, which has
been related to ``antiflow''~\cite{Brachmann:1999xt,Lisa:2000ip} or a ``$3^{\rm
rd}$ component'' of flow~\cite{Csernai:1999nf,Csernai:2011gg}. In the transport
calculations, the emission zone resembles to first order a tilted
ellipsoid~\cite{Lisa:2000ip}, an idealization which is shown in
Fig.~\ref{fig:TiltCartoon}.

\begin{figure}
\centering
\includegraphics[width=0.45\textwidth]{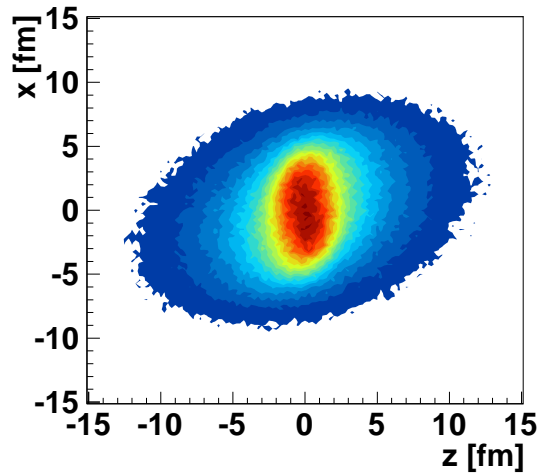}
\caption{Projection of the pion freeze out distribution for Pb+Pb at
$E_{lab}=8$ AGeV, $b=3.4 - 6.8$ fm, $|y|<0.5$ and $p_\perp < 0.4$ GeV.
\label{fig:PiFreeze}}
\end{figure}

We parametrize the tilted freeze out distribution by a three dimensional
 Gaussian ellipsoid in space. In addition we allow that the ellipsoid is rotated by
the angle $\theta_S$ (see Fig. \ref{fig:TiltCartoon}) around the y-axis. This
leads to Eq. \ref{eqn:TiltedFreeze} for the freeze out distribution $f$

\begin{equation}
f(x,y,z) \sim \exp{\left ( - \frac{(x\cos{\theta_S} -
z\sin{\theta_S})^2}{2\sigma_{x'}^2} -\frac{y^2}{2\sigma_y^2}-
\frac{(x\sin{\theta_S} + z\cos{\theta_S})^2}{2\sigma_{z'}^2} \right )}.
\label{eqn:TiltedFreeze}
\end{equation}

In this equation x,y and z are the spatial coordinates, $\theta_S$ is the tilt
angle and $\sigma_{x'}$, $\sigma_y$ and $\sigma_{z'}$ denote the Gaussian
widths of the distribution where the primes on $\sigma_{x'}$ and $\sigma_{z'}$
signify that these correspond to the principal axes of the ellipse instead of
the usual coordinate axes.

Analyzing the freeze out distribution in detail (see Fig. \ref{fig:PiFreeze})
reveals that the system is not characterized by one unique tilt angle, but
exhibits a complex geometry. While the innermost part is almost aligned to the
x-axis ($\theta_S \approx 90^\circ$) the tilt angle is significantly
smaller at the outermost part of the distribution.

\begin{figure}[ht!]
\begin{center}
  \includegraphics[width=0.95\textwidth]{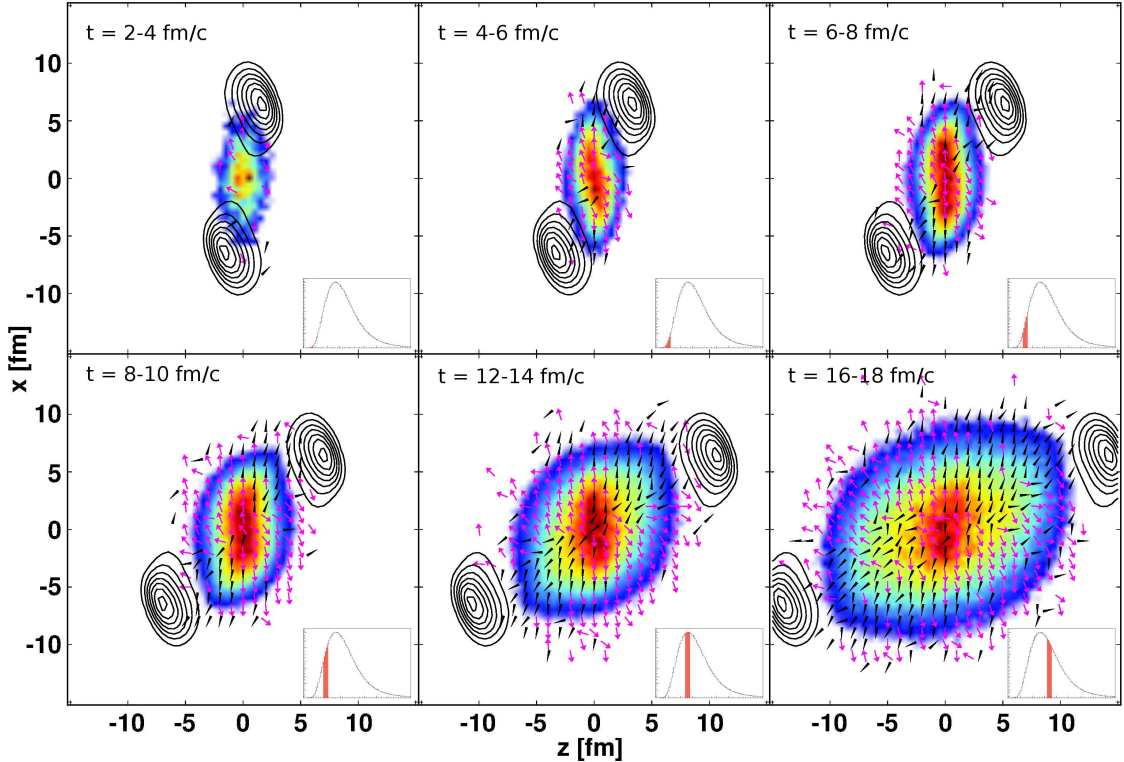}
  \caption{Shape of the freeze out region from pions frozen out at different
times (colored surface). The contour lines depict the position of the spectators
in each timestep. The vector field shows the direction of movement at each
position and time. The black arrowheads contribute to the directed flow while the
magenta arrows contribute to the antiflow. The inlay shows the freeze out
luminosity of pions versus freeze out time. The shaded region in the inlay
highlights the luminosity corresponding to the timestep in the overall
picture. \label{fig:Flow}}
\end{center}
\end{figure}

The source of this structure is seen in Fig. \ref{fig:Flow}. It shows the time
evolution of the pion freeze out distribution (colored surface). The black
contours represent the position of the spectators in each time step while the
vector field depicts direction of the average velocity at each space-time
point. The vector field is split into a directed flow (black arrowheads) and an
antiflow (magenta arrows) component. 
To determine whether a given point in space-time is characterized by flow or
antiflow, the average pion momentum $\overline{\vec{p}}$ is calculated for each
(0.5-fm)$^3$ cell in the $x-z$ plane. Cells with
$\overline{p}_x\cdot\overline{p}_z>0$ (resp. $<0$) are considered to be
dominated by flow (resp. antiflow).

From the time evolution it becomes clear that different angles in the tilt
structure have their origin at different times of the evolution. In the
beginning up to $t\approx4$ fm/c only very few particles are emitted, without
further reinteractions. After that more particles contribute to the freeze out.
From $t\approx4$ fm/c to $t\approx10$ fm/c the emission pattern is dominated by
the fact, that the spectator nucleons shadow the emission of particles in their direction.
This gives rise to an antiflow \cite{Brachmann:1999xt} since most particles
giving a positive contribution to flow are absorbed by the passing target and
projectile nucleons. This automatically leads to a backwards tilt in the freeze
out distribution from early times. 

After $t\approx10$ fm/c the spectators have moved on, so they no longer shadow
the emission. At this point, about 1/3 of pion emission has occurred (see
insets in Fig.~\ref{fig:Flow}). The momentum and spatial anisotropies evolve
differently in the absence of the shadowing influence, leading to a
time-dependent tilt angle. This pattern imprints itself onto the
time-integrated freezeout distribution (Fig.~\ref{fig:PiFreeze}) as a
scale-dependent tilt angle-- the twist. It is the time-integrated freezeout
distribution that is experimentally accessible via HBT measurements, so any
experimental sensitivity to the time evolution of the tilt is through this
twist. 

To underscore the importance of exploring this twist, we point out that even at
the later stages of emission (final panels of Fig.~\ref{fig:Flow}),
the antiflow component in the regions far from the retreating spectators
is as strong as the flow component in the other regions. Thus,
there are {\it two components to antiflow} in these collisions: shadowing
effects in the early stage and preferential spatial expansion along the short
axis of the distribution in the later stage. This latter component is the
analog to the more familiar pressure-gradient-driven elliptic flow in the $x-y$
plane. The interplay between flow and the two sources of antiflow is complex,
presumably energy-dependent, and may be crucial for understanding the details
of $v_1$ measurements.

In most hydrodynamical modeling of elliptic flow, an anisotropic initial state
is generated by some (often ad-hoc) mechanism, and then the system responds
according to pressure gradients, equation of state, etc. In full transport
calculations like UrQMD, there turns out to be at best an approximate factorization
into two stages-- the deposition of energy into the transverse plane at
midrapidity, and the reaction of the system to the initial-state distribution.
However, such a factorization is manifestly impossible when considering
patterns in the $x-z$ plane, where the source is evolving violently even as it
emits particles. Understanding $v_1$ and similar measurements requires a much
more detailed understanding of the space-time evolution of anisotropic
structures.

\begin{figure}
\centering
\includegraphics[width=0.95\textwidth]{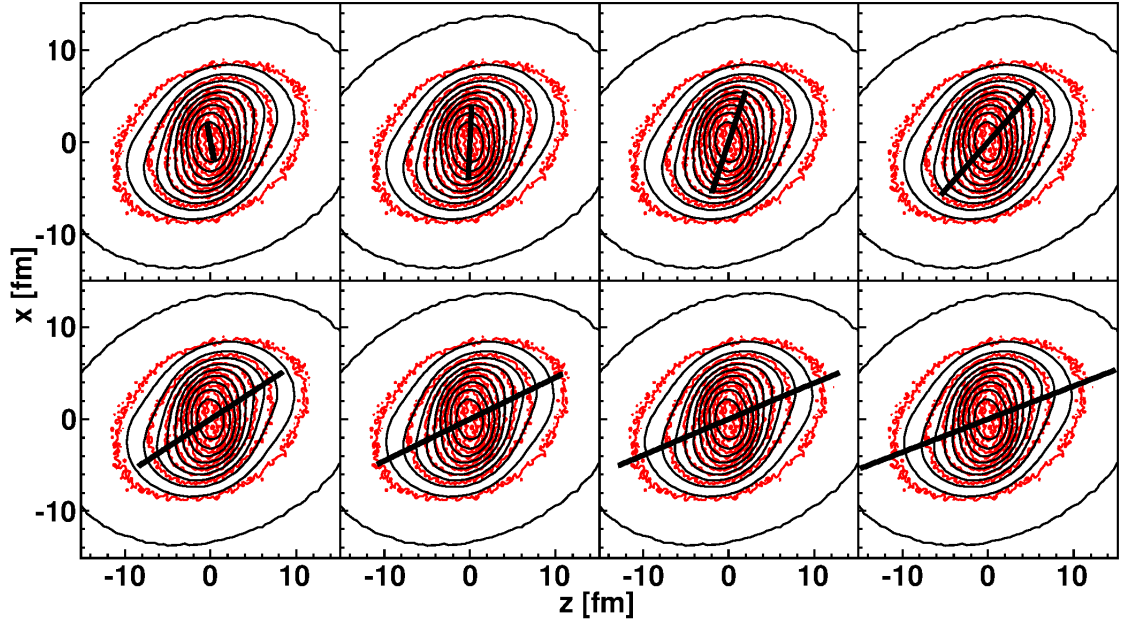}
\caption{Fit to the pion freeze out distribution for Pb+Pb at $E_{lab} = 8 $
GeV, $b= 3.4- 6.8$ fm, $|y|< 0.5$ and $p_\perp <0.4$ GeV. The red contour
is the actual freeze out distribution from UrQMD. The black contour represent the
three dimensional fit to the whole freeze out distribution. The black lines
represent the fit results for the sphere shells. Their angle represents the
fitted angle, while their length represents the radius of the fit
sphere.\label{fig:freezeout}}
\end{figure}

\begin{figure}
\centering
\includegraphics[width=0.5\textwidth]{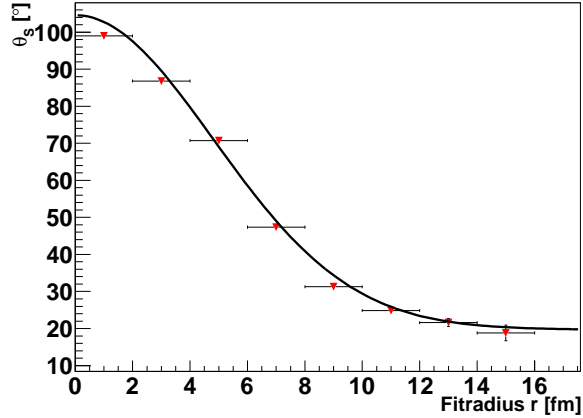}
\caption{Radius dependence of the tilt angle for the fits with rings (red
triangles) and for the full 3D-fit with a radius dependent $\theta_S(r)$ (black
line). The system is Pb+Pb at $E_{lab} = 8$ GeV, $b=3.4 - 6.8$ fm,
$|y| < 0.5$ and $p_\perp < 400 $ MeV. \label{fig:TiltvsRadius}}
\end{figure}

To explore this pattern quantitatively we fit different parts of the distribution
separately, by defining equidistant (2 fm) rings in the x-z plane around
the collision center. Then we perform separate fits for each section of the
freeze out distribution taking into account only the part of the distribution
between two adjacent rings. The y-direction is unrestricted for these fits. The
results of this procedure are shown in Figs. \ref{fig:TiltvsRadius} and
\ref{fig:freezeout}. In Figure \ref{fig:freezeout} the red contours are the
pion freeze out distribution for Pb+Pb at $E_{lab}= 8 $ AGeV, $|y|<0.5$,
$p_\perp^\pi< 400$ MeV. The angle of the black line represents the result for
$\theta_S (r)$ from the separated fitting, while its length describes the outer
limit for the currently applied fit.

In Fig. \ref{fig:TiltvsRadius} the results for the tilt angle of these fits
are presented versus the radius of the fitted segment (red triangles).To
characterize the source effectively we chose to parametrize the $\theta_S (r)$
dependence on $r$ by Eq. \ref{eqn:angle}:

\begin{equation}
\theta_S (r) = \theta_0 + \theta_{Mag} \exp{ \left
(-\frac{r^2}{2\sigma_{twist}^2}\right )}.
\label{eqn:angle}
\end{equation}

Inserting Eq. \ref{eqn:angle} into Eq. \ref{eqn:TiltedFreeze} we gain
a new expression for the freeze out distribution, now with a radius dependent
$\theta_S (r)$. The black contours Fig. \ref{fig:freezeout} represent the
projection of the three dimensional fit of the parametrized freeze out distribution to the
actual freeze out distribution from UrQMD. It describes the overall shape
reasonably well and provides a good description of the tilt angle.
The black line in Fig. \ref{fig:TiltvsRadius} shows the functional dependence of
$\theta_S (r)$ extracted from Eq. \ref{eqn:angle} using the values for the
parameters obtained from the full three dimensional fit. 

\section{Tilt and twist from azimuthally sensitive HBT calculation\label{sec:asHBT}}
The drawback of the procedure described in Section \ref{sec:freeze} is that it
is not possible to measure the spatial freeze out distribution directly in experiment.
If it were possible to measure the twist this would put additional constraints
on many theoretical models. We propose to employ azimuthally sensitive HBT
\cite{Lisa:2000xj,Mount:2010ey,Lisa:2011na,Lisa:2000ip} in a restricted
momentum range to measure the tilt experimentally. Let us briefly outline the
procedure to measure $\theta_S$ via HBT (independent of $r$).

As usual the correlation function is calculated using \cite{Wiedemann:1999qn}
\begin{equation}
  C(\bf{q},\bf{K}) = 1 + \int d^4 x\cos(q\cdot x)d(x,K)
\end{equation}
where $C$ is the correlation function, $q$ is the four-momentum distance of the
correlated particles, $K = (p_1+p_2)/2$ is the pair momentum, $x$ is the
particle separation four-vector and $d$ is the normalized pion freeze out
separation distribution. For the azimuthally sensitive analysis of the HBT
correlations the momentum space is subdivided in several azimuthal sections
around the beam axis. For each of the sections an individual correlation
function is computed. The azimuthal angle of the pair momentum vector
determines in which correlation function each pion pair is counted.

Each of the correlation functions is then fitted separately with 
\begin{equation}
  C(\bf{q},\bf{K}) = 1 + \lambda (\bf{K}) \exp \left [ - \sum_{i,j=o,s,l} q_i
q_j R_{ij}^2 (\bf{K}) \right ] ,
\end{equation}
to obtain the HBT radii $R_{ij}$. For non-central collisions this leads to
oscillating HBT radii with the azimuthal angle $\phi$. Doing a fourier
decomposition it is possible to extract $\theta_S$ for low momentum pairs
using

\begin{equation}
\theta_S = \frac{1}{2} \tan ^{-1} \left (
\frac{-4R_{sl,1}^2}{R_{l,0}^2-R_{s,0}^2+2R_{s,2}^2} \right ) ,
\label{eqn:thetaHBT}
\end{equation}

where in $R_{\nu,\mu}$ the $\mu$ denotes the order of the fourier coefficient,
e.g. $R_{s,2}$ is the second order fourier coefficient of the $R_{s}$
parameter. Details on this method and on finite bin width corrections can be
found in \cite{Mount:2010ey,Wiedemann:1999qn}. While Eq. \ref{eqn:thetaHBT}
allows us to experimentally determine $\theta_S$ it does not give us any
information on the $r$ dependence of $\theta_S (r)$ and it is not clear
how to generalize the derivation of Eq. \ref{eqn:thetaHBT} to a $r$ dependent
$\theta_S$. Thus we resorted to a phenomenological way to determine the twist.

\begin{figure}
\centering
\includegraphics[width=0.75\textwidth]{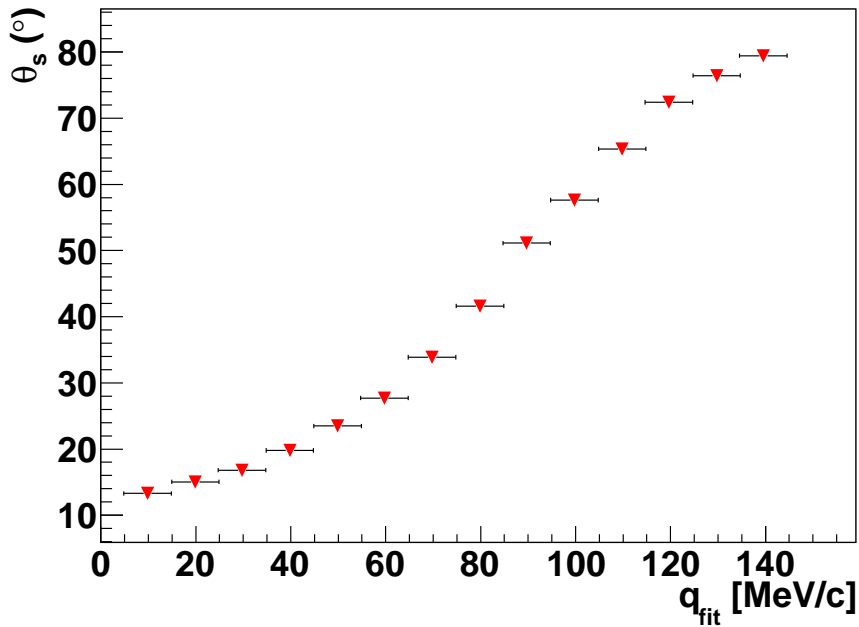}
\caption{Results for $\theta_S(q)$ plotted versus $q_{fit}$, where $q_{fit}$
denotes the $q$ segment of the correlation function that is fitted. The
investigated system is Pb+Pb at $E_{lab}=8$ GeV, $b=3.4-6.8$ fm, $|y| < 0.5$
and $p_\perp < 400$ MeV. \label{fig:HBTTwist}}
\end{figure}

It was already noticed in \cite{Mount:2010ey} that the twist in the freeze out
distribution leads to a rising $\theta_S$ with the fit range of the correlation
functions.This correlation can be attributed to the fact, that larger/smaller
values of $q$ are sensitive to smaller/larger structures of coordinate space.
To get a $\theta_S(r) $ we exploit this behaviour, by applying the same
procedure described in Section \ref{sec:freeze} for the freeze out distribution
in coordinate space, now to the correlation function in momentum space. Namely
we generate correlation functions for eight $45^\circ$-wide bins in $\phi$ to
do the azimuthal HBT analysis. We then define equidistant sphere surfaces in
momentum space around the origin and do the azimuthal HBT analysis needed for
$\theta_S$ in Equation \ref{eqn:thetaHBT} for each sphere shell between two
adjacent sphere surfaces separately. As a result of this procedure we obtain a
$\theta_S(q)$. The result for Pb+Pb, $E_{lab} = 8$ AGeV, $b= 3.4 -6.8 $ fm,
$|y| < 0.6$ and $p_\perp< 0.4$ GeV is shown as red triangles in Fig.
\ref{fig:HBTTwist} versus $q_{fit}$, where $q_{fit}$ is the middle radius of
each sphere shell.

\begin{figure}
\centering
\includegraphics[width=0.75\textwidth]{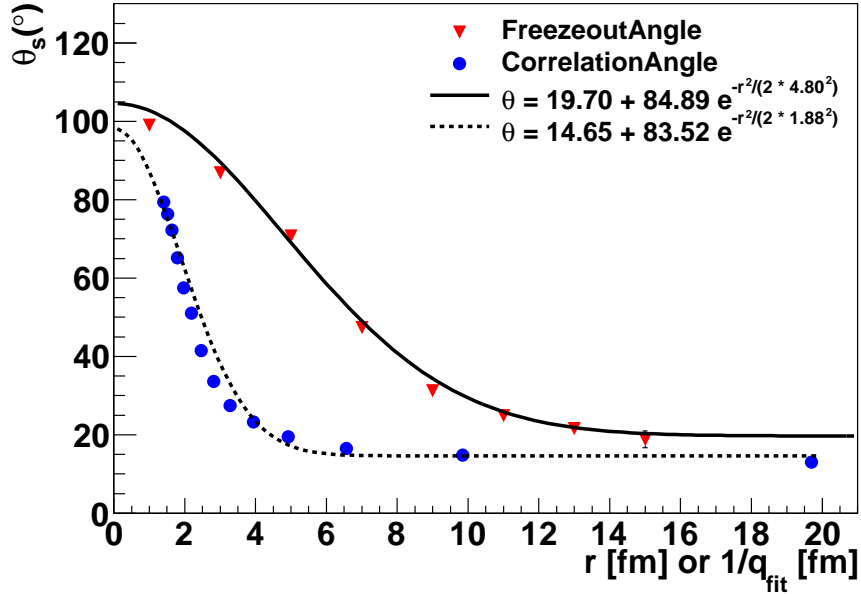}
\caption{$\theta_S$ extracted via fits to the freeze out distribution (red
triangles) and via fits to the HBT correlation functions (blue circles). The
x-axis shows the fit radius $r$ for the freeze out distribution and the inverse
of the momentum fit range $1/q_{fit}$ for the correlation functions. The dashed
black line is a fit of equation \ref{eqn:angle} to $\theta_S$ from the correlation
function. The full black line is also equation \ref{eqn:angle} but using the
parameters of a fully three dimensional fit to the freeze out distribution with
equation \ref{eqn:TiltedFreeze} using the $r$ dependent $\theta_S (r)$.}
\label{fig:RQCombined}
\end{figure}

Indeed, Fig. \ref{fig:TiltvsRadius} and Fig. \ref{fig:HBTTwist} do bear a
striking similarity if one keeps in mind, that larger $q$ values mean
sensitivity to smaller regions of homogeneity and vice versa. In Fig.
\ref{fig:RQCombined} we compare the results of both methods (see Figs.
\ref{fig:TiltvsRadius} and \ref{fig:HBTTwist}) in a single figure. The symbols
show $\theta_{S}(q)$ versus 1/q (blue circles) and $\theta_{S}(r)$ versus r
(red triangles).  It is clear that a simple $r \sim 1/q$, as done for the plot
reflects the relation between the spatial extension of the source and the
region of homogeneity only qualitatively (there might be some other
proportionality factor, that depends on the flow and temperature). Nevertheless
it clearly indicates the momentum bin differential azimuthally sensitive HBT
allows to capture the complicated source structure. Let us now compare the
calculations to the expectations obtained from Eqs. \ref{eqn:TiltedFreeze} and
\ref{eqn:angle}. The dashed black line in Fig. \ref{fig:RQCombined} is a fit of
equation \ref{eqn:angle} to $\theta_S (1/q)$. The description of the
theoretical data points is very good and leads to $\theta_{Mag}$ and
$\theta_{0}$ of both methods being similar to each other.

\section{Summary and discussion}
The analysis of the spatial pion freeze out distribution of a non-central lead
lead collision from UrQMD features a scale-dependent tilt, or ``twist''. The
twist originates from antiflow and shadowing of pion emission at early times
and the absence of shadowing at later times. To make the spatial twist
experimentally accessible we have employed an azimuthally sensitive HBT
analysis that allows to measure the tilt angle. Using the fact that pairs with
small momentum difference are sensitive to large space-time structures and vice
versa we calculated the tilt angle on different scales. The analysis shows that
this procedure provides a qualitatively accurate picture of the radius
dependence of the tilted freeze out distribution. This analysis enables us for
the first time to use HBT correlations to disentangle the geometry of the
source from different times up to a certain point. We conclude that the twist
structure is in principle accessible by experimental HBT analysis and may allow
to gain complementary insights into the early emission stages of the reaction.

\section*{Acknowledgements}
This work was supported by the Hessian LOEWE initiative through Helmholtz
International Center for FAIR (HIC for FAIR). The Frankfurt Center for
Scientific Computing(CSC) provided the computational resources. G.G. thanks the
Helmholtz Research School for Quark Matter Studies (H-QM) for support.


\section*{References}

\begin{thebibliography}{36}

\bibitem{Lisa:2005dd} 
  M.~A.~Lisa, S.~Pratt, R.~Soltz and U.~Wiedemann,
  Ann.\ Rev.\ Nucl.\ Part.\ Sci.\  {\bf 55}, 357 (2005)
  [nucl-ex/0505014].

\bibitem{Lisa:2000xj} 
  M.~A.~Lisa {\it et al.}  [E895 Collaboration],
  Phys.\ Lett.\ B {\bf 496}, 1 (2000)
  [nucl-ex/0007022].

\bibitem{Adams:2003ra} 
  J.~Adams {\it et al.}  [STAR Collaboration],
  Phys.\ Rev.\ Lett.\  {\bf 93}, 012301 (2004)
  [nucl-ex/0312009].

\bibitem{Adamova:2008hs} 
  D.~Adamova {\it et al.}  [CERES Collaboration],
  Phys.\ Rev.\ C {\bf 78}, 064901 (2008)
  [arXiv:0805.2484 [nucl-ex]].

\bibitem{Aggarwal:2010cw} 
  M.~M.~Aggarwal {\it et al.}  [STAR Collaboration],
  arXiv:1007.2613 [nucl-ex].

\bibitem{Anson:2011ik} 
  C.~Anson [STAR Collaboration],
  J.\ Phys.\ G G {\bf 38}, 124148 (2011)
  [arXiv:1107.1527 [nucl-ex]].

\bibitem{Bleicher:1999xi}
  M.~Bleicher {\it et al.},
  J.\ Phys.\ G {\bf 25}, 1859 (1999)
  [arXiv:hep-ph/9909407].

\bibitem{Bass:1998ca}
  S.~A.~Bass {\it et al.},
  Prog.\ Part.\ Nucl.\ Phys.\  {\bf 41}, 255 (1998)
  [Prog.\ Part.\ Nucl.\ Phys.\  {\bf 41}, 225 (1998)]
  [arXiv:nucl-th/9803035].

\bibitem{Li:2006gp} 
  Q.~Li, M.~Bleicher and H.~Stoecker,
  Phys.\ Rev.\ C {\bf 73}, 064908 (2006)
  [nucl-th/0602032].
\bibitem{Li:2007yd} 
  Q.~Li, M.~Bleicher and H.~Stocker,
  Phys.\ Lett.\ B {\bf 659}, 525 (2008)
  [arXiv:0709.1409 [nucl-th]].
 
\bibitem{Li:2008qm} 
  Q.~-f.~Li, J.~Steinheimer, H.~Petersen, M.~Bleicher and H.~Stocker,
  Phys.\ Lett.\ B {\bf 674}, 111 (2009)
  [arXiv:0812.0375 [nucl-th]].

\bibitem{Li:2012ta} 
  Q.~Li, G.~Graf and M.~Bleicher,
  Phys.\ Rev.\ C {\bf 85}, 034908 (2012)
  [arXiv:1203.4104 [nucl-th]].

\bibitem{Brachmann:1999xt} 
  J.~Brachmann, S.~Soff, A.~Dumitru, H.~Stoecker, J.~A.~Maruhn, W.~Greiner, L.~V.~Bravina and D.~H.~Rischke,
  Phys.\ Rev.\ C {\bf 61}, 024909 (2000)
  [nucl-th/9908010].

\bibitem{Mount:2010ey} 
  E.~Mount, G.~Graef, M.~Mitrovski, M.~Bleicher and M.~A.~Lisa,
  Phys.\ Rev.\ C {\bf 84}, 014908 (2011)
  [arXiv:1012.5941 [nucl-th]].

\bibitem{Lisa:2011na} 
  M.~A.~Lisa, E.~Frodermann, G.~Graef, M.~Mitrovski, E.~Mount, H.~Petersen and M.~Bleicher,
  New J.\ Phys.\  {\bf 13}, 065006 (2011)
  [arXiv:1104.5267 [nucl-th]].

\bibitem{Lisa:2000ip} 
  M.~A.~Lisa, U.~W.~Heinz and U.~A.~Wiedemann,
  Phys.\ Lett.\ B {\bf 489}, 287 (2000)
  [nucl-th/0003022].

\bibitem{Wiedemann:1999qn} 
  U.~A.~Wiedemann and U.~W.~Heinz,
  Phys.\ Rept.\  {\bf 319}, 145 (1999)
  [nucl-th/9901094].

\bibitem{Ollitrault:1992bk} 
  J.~-Y.~Ollitrault,
  Phys.\ Rev.\ D {\bf 46}, 229 (1992).

\bibitem{Csernai:1999nf} 
  L.~P.~Csernai and D.~Rohrich,
  Phys.\ Lett.\ B {\bf 458}, 454 (1999)
  [nucl-th/9908034].

\bibitem{Magas:2000cm} 
  V.~K.~Magas, L.~P.~Csernai and D.~D.~Strottman,
  hep-ph/0101125.

\bibitem{Csernai:2011gg} 
  L.~P.~Csernai, V.~K.~Magas, H.~Stocker and D.~D.~Strottman,
  Phys.\ Rev.\ C {\bf 84}, 024914 (2011)
  [arXiv:1101.3451 [nucl-th]].

\end{thebibliography}
\end{document}